\documentclass[12pt]{iopart}
\input epsf
\usepackage{iopams}
\usepackage{bm}

\newcommand{\apj}{\textit{Astrophys.\ J.} }
\newcommand{\aanda}{\textit{Astron.\ Astrophys.} }

\newcommand{\pre}{\textit{Phys.\ Rev.} E }
\newcommand{\pra}{\textit{Phys.\ Rev.} A }

\newcommand{\gcc}{\mathrm{~g~cm}^{-3}}

\begin{document}

\article[Strongly magnetized plasmas of neutron stars]{Strongly 
Coupled Coulomb Systems - 2005}{Nonideal strongly magnetized plasmas of neutron stars and
their electromagnetic radiation}

\author{A Y Potekhin$^1$, G Chabrier$^2$, D Lai$^3$, W C G Ho$^4$
 and M van Adelsberg$^3$}

\address{$^1$ Ioffe Physico-Technical Institute, St Petersburg, Russia}
\address{$^2$ Ecole Normale Sup\'erieure de Lyon, C.R.A.L.(UMR 5574 CNRS), France}
\address{$^3$ Center for Radiophysics and Space Research, Department of Astronomy,
Cornell University, Ithaca, NY 14853, USA}
\address{$^4$ Hubble Fellow;
MIT Kavli Institute for Astrophysics and Space Research,
77 Massachusetts Ave, 37-582F, Cambridge, MA 02139 USA}
\eads{\mailto{palex@astro.ioffe.ru}}

\begin{abstract}
We study the equation of state, polarization
and radiation properties
for nonideal, strongly magnetized plasmas
which compose outer envelopes of magnetic neutron stars. Detailed
calculations are performed for partially ionized
 hydrogen atmospheres and for condensed 
 hydrogen or iron surfaces of these stars.
\end{abstract}

\submitto{\JPA}

\section{Introduction}

Neutron stars can be considered as natural laboratories for
studying the properties of matter under extreme physical
conditions (see, e.g., \cite{ST83}). 
In their cores the density $\rho$ may exceed
$10^{15}$ g cm$^{-3}$ and the temperature $T$
lies typically between $10^7$ and $10^9$~K, but their thermal emission depends on the
properties of the outer envelopes with lower $\rho$ and $T$.
The temperature
decreases in
the heat-blanketing envelopes
(at $\rho \lesssim 10^8$ g cm$^{-3}$) to 
$T\sim T_\mathrm{eff}$ near the radiative surface 
(at $\rho \sim10^{-3}$--$10^6$ g cm$^{-3}$, depending
on the stellar parameters), where $T_\mathrm{eff}$ 
is the \emph{effective surface temperature} related 
to the thermal flux through the Stefan's law.
In these envelopes,
the magnetic field strength $B$ may reach $10^{15}$ G; most often
$B\sim 10^{11} - 10^{13}$ G. 
To calculate spectra of neutron-star emission 
that can be compared with the observations, one
should take into account nonideality of the plasma in
the envelopes
and the effects of strong magnetic fields on the
properties of the plasma and electromagnetic radiation. 

In the cases
where $T_\mathrm{eff}\sim\mbox{a few}\times10^5$--$10^6$~K
(characteristic of
middle-aged neutron stars, i.e., of age $\sim10^4$--$10^6$ yr)
and $B\lesssim10^{13}$~G,
the spectrum of thermal emission from a neutron star forms in
an stellar \emph{atmosphere}. 
In a strong magnetic field, the state of matter
can change significantly (for review, see, e.g., \cite{Lai01}). 
For instance, at $T_\mathrm{eff}\sim10^6$~K, 
an atmosphere composed of hydrogen might be treated
as fully ionized if the magnetic field were zero, but
there can be a significant fraction 
of bound H atoms, if $B\gtrsim10^{11}$~G. 
At lower $T$ (characteristic of older stars)
and sufficiently high $B$, \emph{magnetic
condensation} may occur, resulting in formation of a solid
or liquid metallic surface,
composed of a strongly coupled Coulomb plasma
and covered by a thin gaseous atmosphere.

Although the possible importance of these effects
has been realized long ago, they 
were not included in 
neutron-star atmosphere models until recently
(e.g., \cite{Pavlov95} and references therein).
Early considerations of partial ionization
in the magnetized neutron-star atmospheres 
(e.g., \cite{Miller92,RRM}; also reviewed briefly in \cite{ZP02}) 
were
not very reliable because of oversimplified treatments of
atomic physics and nonideal plasma effects in strong magnetic fields.
However, in the last decade the studies of dense nonideal plasmas
in the strong magnetic fields made a spectacular progress,
that now allows one to calculate the spectrum and polarization
of electromagnetic radiation formed in an extended
partially ionized hydrogen atmosphere or emitted from 
a condensed metallic surface of a magnetic neutron star.
In this paper, we briefly review these achievements.

\section{Equation of state and opacities of partially ionized atmospheres
with strong magnetic fields}

Motion of 
electrons and ions in a magnetic field 
is quantized in the Landau orbitals. In neutron-star atmospheres,
the field is usually \emph{strongly quantizing} for the electrons
(see, e.g., companion paper \cite{CSP} for basic
definitions). When $B\gg10^9$~G,
the properties of atoms and molecules change drastically
\cite{Lai01}. In particular, the center-of-mass
motion of an atom is coupled to the internal degrees of freedom.
This effect is important for both the equation of state (EOS)
and radiative opacities of the plasma.
In this section, we briefly review these effects for the case
of the hydrogen plasma, the only one that has been studied
in detail up to now. This case is quite realistic,
because even a small amount of hydrogen
 on the surface of the neutron star is sufficient to fill
 the entire atmosphere.
For instance, for a ``canonical'' neutron star 
(with mass $M=1.4$ solar masses
and radius $R=10$ km), the total mass of the hydrogen atmosphere
is $\sim10^{12}\rho T_6$~g,
where $\rho$ is the bottom atmosphere density in g cm$^{-3}$ and
$T_6=T_\mathrm{eff}/10^6$~K.
This mass of hydrogen can be accreted from 
the interstellar medium or from a nebula
surrounding the star, or produced in spallation of heavy
nuclei by highly energetic plasma discharges
in the polar-cap regions of a pulsar.

\subsection{Ionization equilibrium and equation of state}

The EOS for partially ionized hydrogen 
at $10^{11}$~G$\lesssim B \lesssim 10^{13}$~G
was calculated and discussed in \cite{PCS99};
recently it was extended 
to the case $10^{13}$~G$\lesssim B \lesssim 10^{15}$~G
\cite{PC04} and
tabulated \cite{PC04,PC03}. The treatment is
based on the free energy minimization. 
The free energy model is in essence a generalization of the 
nonmagnetic model of Saumon and Chabrier \cite{SC91,SC92}.
We consider a plasma composed of $N_\mathrm{p}$ protons, 
$N_\mathrm{e}$ electrons, $N_\mathrm{H}$ hydrogen atoms, and
$N_\mathrm{mol}$ molecules in a volume $V$, the number densities being
$n_j\equiv N_j/V$. The Helmholtz free energy is written as
\begin{equation}
   F = F_\mathrm{id}^\mathrm{e} + F_\mathrm{id}^\mathrm{p} 
      + F_\mathrm{id}^\mathrm{neu}
       + F_\mathrm{ex}^\mathrm{C} + F_\mathrm{ex}^\mathrm{neu},
\label{Fren}
\end{equation}
where $F_\mathrm{id}^\mathrm{e}$, $F_\mathrm{id}^\mathrm{p}$, and
$F_\mathrm{id}^\mathrm{neu}$ are the free energies of ideal gases
of the electrons, protons, and neutral species, respectively, 
$F_\mathrm{ex}^\mathrm{C}$ takes into account the Coulomb plasma
nonideality, and $F_\mathrm{ex}^\mathrm{neu}$ is the nonideal
contribution which arises from interactions of bound species with
one another and with the electrons and protons. 
Ionization equilibrium is given by minimization of $F$ with
respect to particle numbers under the stoichiometric constraints,
keeping $V$ and
the total number of protons (free and bound)
constant.
The formulae for each term in (\ref{Fren}) are given in
\cite{PCS99,PC03}. 
The employed minimization technique is similar to
the one presented in \cite{P96} for the nonmagnetic case.

\begin{figure}
\begin{flushright}
\epsfxsize=140mm
\epsfbox{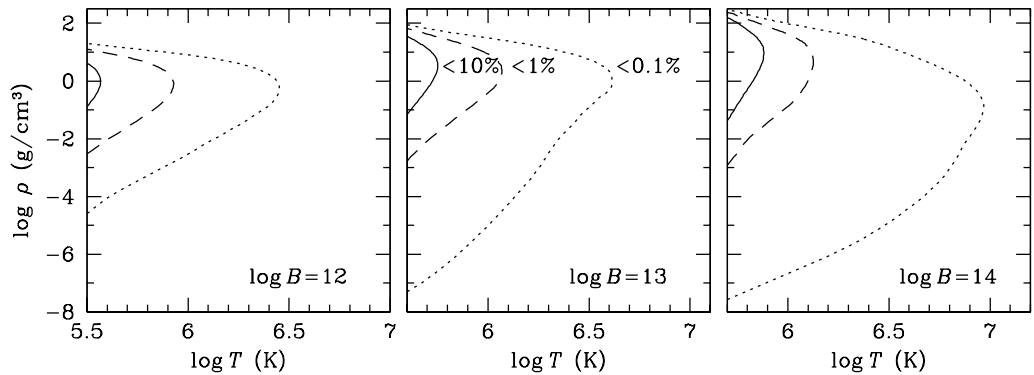}
\end{flushright}
\caption{Domains of partial ionization in the $\rho-T$ plane
for $B=10^{12}$, $10^{13}$ and $10^{14}$~G. 
The contours delimit the domains where the atomic fraction
$x_\mathrm{H}<0.1$\% (to the right of the dotted lines), 
0.1\%$<x_\mathrm{H}<1$\% (between the dashed and dotted lines), 
1\%$<x_\mathrm{H}<10$\% (between the solid and dashed lines)
or $x_\mathrm{H}>10$\% (to the left of the solid lines).
\label{phase}}
\end{figure}

Once the free energy is obtained, its derivatives 
over $\rho$ and $T$ and their combinations provide
the other thermodynamic functions. These functions
are  affected by the strong magnetic field.
In the domain of strongly quantizing field ($\rho<\rho_B$,
see \cite{CSP}), the electron degeneracy is reduced
and the EOS becomes softer than it would be
at $B=0$. If the field is weakly quantizing ($\rho\gtrsim\rho_B$),
the second-order thermodynamic functions
 oscillate with increasing $\rho$ around their
 field-free values (for details, see \cite{PCS99,PC03}). 

Figure \ref{phase} shows the domains of partial ionization 
in the $\rho-T$ plane
for different values of $B$. The higher $B$, the greater $T$ at which the
bound species are important. The calculated atomic fractions 
$x_\mathrm{H}$
are needed to obtain the radiative opacities in the atmosphere
(see \S\ref{sect:opacity} below).

Our model becomes less reliable 
at relatively low $T$
where molecular chains H$_n$ can be important \cite{Lai01}. Generally,
this occurs within the $\rho-T$ domain where
$x_\mathrm{H}\gtrsim0.1$ (i.e., to the left of the solid lines
in Fig.~\ref{phase}).

\subsection{Radiative opacities, polarization and spectra}
\label{sect:opacity}

A magnetic field affects the properties of
electromagnetic radiation in plasmas
 (e.g., \cite{Ginzburg}).
At photon energies $\hbar\omega$ much higher than
$
   \hbar\omega_\mathrm{pl} =
        ({4\pi\hbar^2 e^2 n_\mathrm{e} 
                      / {m_e}} )^{1/2}
                      \approx 28.7\,\rho^{1/2}  \mathrm{~eV},
$
where 
 $\omega_\mathrm{pl}$ is the electron plasma frequency and 
$\rho$ is the density in $\gcc$,
radiation propagates in the form of two so-called \emph{normal modes}.
These modes have different polarization vectors
$\bm{e}_j$
and different absorption and scattering
coefficients, which depend on the angle $\theta_B$
between the radiation propagation direction and the magnetic field.
The two modes interact with each other via scattering.

\begin{figure}
\begin{center}
\epsfxsize=105mm
\epsfbox{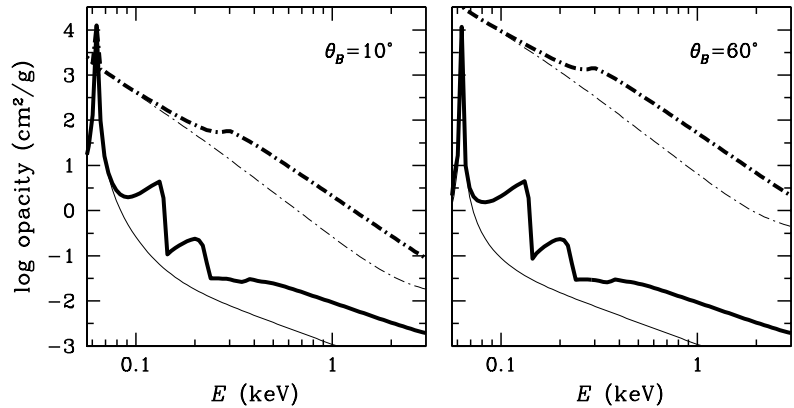}
\end{center}
\caption{Radiative opacities of the plasma versus photon energy
at $B=10^{13}$~G, $\rho=1$ g cm$^{-3}$ and $T=10^6$~K,
for two values of the angle $\theta_B$
between the propagation direction and the magnetic field
(left: $\theta_B=10^\circ$; right: $\theta_B=60^\circ$),
for the ordinary (chain lines) and extraordinary (solid lines)
normal mode of radiation. The case of partially ionized
hydrogen plasma (thick lines) is compared with the
approximation of fully ionized plasma (thin lines).
\label{opacity}}
\end{figure}

In a partially ionized atmosphere, the opacity 
for each normal mode is contributed
by electrons, ions, and bound species. 
The scattering cross section includes contributions from the
electrons and protons: 
$
   \sigma_\alpha^\mathrm{s} = \sigma_\alpha^\mathrm{s,e}
              + \sigma_\alpha^\mathrm{s,p}$
(the subscript $\alpha=0,\pm1$ denotes
a basic photon polarization 
 -- longitudinal, right or left circular -- 
 with respect to the magnetic field
direction).
The absorption cross section $\sigma_\alpha^\mathrm{a}$
includes contributions from absorption by plasma electrons and protons
(mainly by free-free transitions due to the electron-proton collisions,
 $\sigma_\alpha^\mathrm{ff}$),
transitions between discrete states of an atom
(bound-bound absorption, $\sigma_\alpha^\mathrm{bb}$) and
photoionization (bound-free, $\sigma_\alpha^\mathrm{bf}$).
Thus, for the hydrogen atmosphere, 
we can write
$
   \sigma_\alpha^\mathrm{a} \approx 
   x_\mathrm{H} (\sigma_\alpha^\mathrm{bb}+\sigma_\alpha^\mathrm{bf})
   + (1-x_\mathrm{H})\,\sigma_\alpha^\mathrm{ff} .
$
For the hydrogen in a strong
magnetic field, the cross sections $\sigma_\alpha^\mathrm{bb}$
were studied in \cite{PP95},
$\sigma_\alpha^\mathrm{bf}$ in \cite{PP97},
and $\sigma_\alpha^\mathrm{ff}$ in \cite{PC03}.

The bound species affect
the dielectric tensor of the medium and hence the 
polarization properties of the normal modes.
These effects were studied in \cite{KK},
with use of the Kramers-Kronig relation
between the imaginary and real parts of the plasma polarizability.
The full account of the coupling of the
quantum mechanical structure of the atoms to their center-of-mass
motion across the magnetic field proved to be crucial for the correct
evaluation of the plasma polarization properties and opacities.

An example of the opacities is shown 
in Fig.~\ref{opacity}. In this case, 
the ionization degree is high ($x_\mathrm{H}=0.016$),
but the account of the bound species is still
important: in the photon energy ranges corresponding to
the bound-bound and bound-free atomic transitions,
the model of a fully ionized plasma underestimates
the opacities by a factor of $\sim 10$.

\begin{figure}
\begin{flushright}
\epsfxsize=150mm
\epsfbox{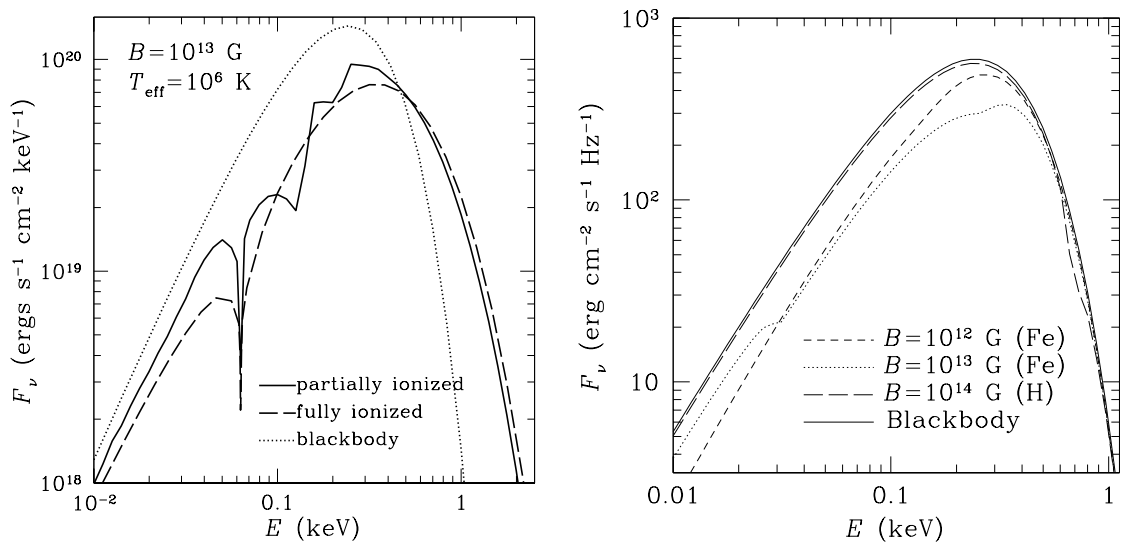}
\end{flushright}
\caption{Spectral flux as a function of 
photon energy $E$.
\emph{Left}: The case of a  partially ionized
hydrogen atmosphere (solid line)
at $B=10^{13}$~G (field normal to the surface)
and $T_\mathrm{eff}=10^6$~K
is compared with the
approximation of fully ionized plasma (dashes)
and with the blackbody spectrum (dots).
\emph{Right}: 
The cases of condensed Fe 
($B=10^{12}$ G, short dashes; $10^{13}$ G, dots) and H ($B=10^{14}$ G,
long dashes)
surfaces, at temperature $T = 10^6$ K, compared with
the blackbody spectrum (solid line).
\label{spectrum}}
\end{figure}

These results are used in modeling the spectra of radiation 
from the neutron-star atmospheres \cite{KK,Ho-ea03}. An example shown
on the left panel of Fig.~\ref{spectrum} clearly demonstrates the importance of 
the partial-ionization effects.

\section{Radiation from the condensed surface
in a strong magnetic field}

The notion that an isolated magnetic neutron star has a 
condensed surface was first put forward by Ruderman 
\cite{Ruderman71}, who considered the Fe surface.
 Lai \& Salpeter \cite{LS97} (see also \cite{Lai01})
 studied the phase diagram of strongly magnetized hydrogen and
showed that, if 
$T_\mathrm{eff}$ is below a critical value (which is a function of 
$B$), the atmosphere can undergo a phase
transition into a condensed state.
The thermal emission from the magnetized Fe surface was  
studied in \cite{Brinkmann,Turolla-ea,perez-azorin} and most thoroughly
in paper \cite{surfem}. 
For a smooth condensed surface, the overall emission is 
reduced from the 
blackbody by less than a factor of 2,
whereas for a 
rough surface, the reduction can be negligible 
(see \cite{surfem}). The spectrum exhibits 
modest deviation from blackbody across a wide energy range, 
and shows mild absorption features associated with the ion 
cyclotron frequency and the electron plasma frequency in the condensed 
matter. 
Examples of the spectrum
for different models of the surface are shown
on the right panel of Fig.~\ref{spectrum}.

\section{Concluding remarks}

We have briefly reviewed the main effects of
strong magnetic fields on the EOS, opacities
and properties of electromagnetic radiation
in the surface layers of neutron stars.
In order to calculate realistic spectra
of thermal radiation from neutron stars,
one must carefully take these effects
into account.
We expect that comparison of the calculated
models with observations will 
help to improve the constraints on neutron star
parameters (for example, their effective temperatures,
see \cite{CSP}) and thus
provide
a powerful tool for testing the theories
of superdense matter in neutron star interiors.

\ack
The work of GC was partially supported
by the CNRS French-Russian grant PICS 3202.
The work of AYP was partially supported by
the RLSS grant 1115.2003.2 
and the RFBR grants 05-02-16245, 03-07-90200, and 05-02-22003. 
WCGH is supported by NASA through Hubble Fellowship grant
HF-01161.01-A awarded by the STScI, which is
operated by the AURA, Inc, for NASA, under contract NAS~5-26555.

\end{document}